\documentclass[conference]{IEEEtran}
\IEEEoverridecommandlockouts
\usepackage{cite}
\usepackage{amsmath,amssymb,amsfonts}
\usepackage{algorithmic}
\usepackage{graphicx}
\usepackage{textcomp}
\usepackage{xcolor}
\def\BibTeX{{\rm B\kern-.05em{\sc i\kern-.025em b}\kern-.08em
    T\kern-.1667em\lower.7ex\hbox{E}\kern-.125emX}}
\begin{document}

\title{Edge-PRUNE: Flexible Distributed Deep Learning Inference\\
\thanks{This research was partially funded by the Academy of Finland project SPHERE-DNA (grants  345681 and 345683).
}
}

\author{\IEEEauthorblockN{Jani Boutellier}
\IEEEauthorblockA{\textit{School of Technology and Innovations} \\
\textit{University of Vaasa}\\
Vaasa, Finland \\
jani.boutellier@uwasa.fi}
\and
\IEEEauthorblockN{Bo Tan}
\IEEEauthorblockA{\textit{ITC Faculty} \\
\textit{Tampere University}\\
Tampere, Finland \\
bo.tan@tuni.fi}
\and
\IEEEauthorblockN{Jari Nurmi}
\IEEEauthorblockA{\textit{ITC Faculty} \\
\textit{Tampere University}\\
Tampere, Finland \\
jari.nurmi@tuni.fi}
}

\maketitle

\begin{abstract}
Collaborative deep learning inference between low-resource endpoint devices and edge servers has received significant research interest in the last few years. Such computation partitioning can help reducing endpoint device energy consumption and improve latency, but equally importantly also contributes to privacy-preserving of sensitive data. This paper describes Edge-PRUNE, a flexible but light-weight computation framework for distributing machine learning inference between edge servers and one or more client devices. Compared to previous approaches, Edge-PRUNE is based on a formal dataflow computing model, and is agnostic towards machine learning training frameworks, offering at the same time wide support for leveraging deep learning accelerators such as embedded GPUs. The experimental section of the paper demonstrates the use and performance of Edge-PRUNE by image classification and object tracking applications on two heterogeneous endpoint devices and an edge server, over wireless and physical connections. Endpoint device inference time for SSD-Mobilenet based object tracking, for example, is accelerated 5.8$\times$ by collaborative inference.
\end{abstract}


\section{Introduction}

\noindent In the last few years there has been an increasing interest both in academia and in the industry to find ways to move the computation effort of deep learning inference from centralized clouds and servers closer to the network edge and to endpoint devices (mobiles, smart cameras, \textit{etc.}). For widely deployed deep learning applications, such as speech recognition \cite{zhang2021tiny}, performing inference on the endpoint device reduces network bandwidth usage, latency, and need for exposing potentially sensitive data to the network. On the other hand, deep learning inference requires the endpoint device to perform a significant amount of computations, which is potentially not feasible for very low-end mobile devices.

Various approaches for improving the efficiency of deep learning inference have been proposed, ranging from hardware accelerators (\textit{e.g.}, \cite{skillman2020technical}) to computational optimizations \cite{zhu2017prune, rastegari2016xnor, jaderberg2014speeding}. Orthogonal to these techniques that speed up inference on the endpoint device, also various \textit{collaborative inference} approaches have been proposed \cite{li2018edge, kang2017neurosurgeon, laskaridis2020spinn, jeong2018ionn}. In collaborative inference, the computational load is partitioned over a network connection between endpoint devices and cloud/server resources. The seminal work Neurosurgeon \cite{kang2017neurosurgeon} originally proposed splitting the chain-like layer structure of a deep neural network (DNN) such that the inference of the early layers is performed by the endpoint device, whereas the inference of the later layers is performed in the cloud. Determining the optimal cut-off point then depends on the DNN architecture, computational resources of the endpoint device, and network bandwidth.

\begin{figure}
\centering
\includegraphics[width=\linewidth]{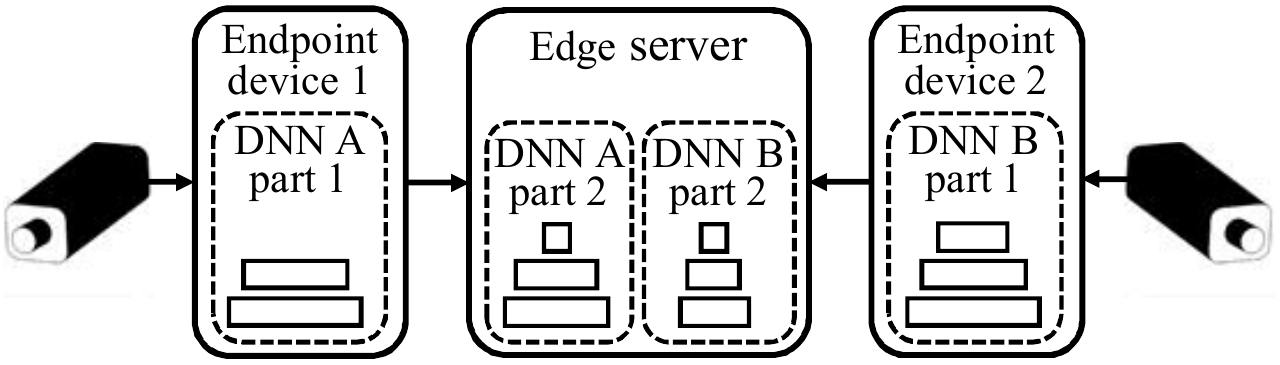}
\caption{An example of a distributed image recognition system: heterogeneous endpoint devices 1 and 2 are both connected to a camera and perform image recognition using different DNN architectures A and B. The inference of both DNNs has been partitioned across the endpoint devices and the edge server.}
\label{fig:overview}
\end{figure}

Distributed deep learning has in the last few years gained attention also from the data privacy point of view, since transmission of training or inference data over the wireless network exposes the content to various attacks \cite{liu2020privacy}. Compared to distributed training of deep learning models, distributed deep learning inference is somewhat less susceptible to data privacy compromising attacks, since the model parameters do not depend on the inference input, and the inference input does not follow any statistical distribution \cite{he2019model}. Nevertheless, by means of selected black-box, white-box and query-free attacks \cite{he2019model}, an adversary can still recover the inference input from intermediate data that has been produced by the early DNN layers on the endpoint device, and has been transmitted over the network to the cloud for completing the inference. \textit{One of the most important defenses against such attacks is to perform the inference of as many DNN layers as possible on the endpoint device}, because recovering the input becomes increasingly harder as it passes through DNN layers \cite{he2019model}.

In order to perform as much of the inference on the endpoint device as possible, it is of highest importance to be able to leverage the performance of computation accelerators (\textit{e.g.} \cite{skillman2020technical}). This paper proposes \textit{Edge-PRUNE}, a flexible framework for distributed deep learning inference with wide support for various inference accelerators, such as GPUs, CPU SIMD instructions, \textit{etc.} The key features of Edge-PRUNE include:
\begin{itemize}
\item A solid dataflow model of computation,
\item Framework\footnote{Available at https://gitlab.com/jboutell/vprf/-/tree/edge-prune} and tools for synthesizing code to heterogeneous endpoint and server devices,
\item An efficient runtime that takes care of client-server communication, inter-process communication and hardware accelerator interfacing.
\end{itemize}
Edge-PRUNE is agnostic towards machine learning training frameworks (such as TensorFlow or PyTorch), but allows leveraging DNN acceleration libraries such as Intel oneDNN and ARM CL, if needed. The generic nature of Edge-PRUNE also enables heterogeneous collaborative inference scenarios (see Fig.~\ref{fig:overview}), where the endpoint devices and the edge server can run different DNN architectures and use hardware accelerators from various vendors.

The rest of the paper is organized as follows: Section~\ref{sec:related} briefly reviews relevant related works, Section~\ref{sec:proposed} presents the Edge-PRUNE framework, Section~\ref{sec:experiments} shows experimental evaluation of the proposed work, Section~\ref{sec:discussion} discusses the results, and Section~\ref{sec:conclusions} concludes the paper.

\section{Related work}
\label{sec:related}

\noindent Recently, distributed machine learning inference has received a lot of attention. The pioneering work Neurosurgeon \cite{kang2017neurosurgeon} proposed a prediction-based scheduler on top of Caffe, for automatic partitioning of DNNs between mobile devices and data centers. Simultaneously, the DDNN framework \cite{teerapittayanon2017distributed} proposed homogeneously distributed machine learning with \textit{early exits} across cloud, edge and multiple endpoint devices. Recently, the DDNN framework has been implemented within the Adaptive Computing Framework (ACF) \cite{tu2021designing}. In contrast to endpoint-server collaborative inference, several works \cite{mao2017modnn,  zhao2018deepthings, gao2021edgesp} have proposed the distribution of DNN inference across multiple endpoint devices.

Edgent \cite{li2018edge} (and its successor Boomerang \cite{zeng2019boomerang}) continued in the vein of Neurosurgeon, proposing DNN \textit{right-sizing} through the use of early exits (similar to DDNN \cite{teerapittayanon2017distributed}), with an implementation based on the Chainer deep learning framework. In contrast, IONN \cite{jeong2018ionn} proposed an incremental offloading scheme, where a client device uploads partitions of a DNN model to a server for distributed inference; similar to Neurosurgeon, IONN is Caffe-based. JointDNN \cite{eshratifar2019jointdnn} formulated the partitioning of computations as a graph shortest path problem, also paying attention to autoencoder and generative model type DNNs, where output size grows towards the last layers. JALAD \cite{li2018jalad} proposed an optimization framework for distributing DNN computations between edge and cloud, considering the possibility of feature compression, accuracy and latency optimization. 

DADS \cite{hu2019dynamic} proposed the directed acyclic graph (DAG) based edge-cloud DNN inference model (ECDI) that can be used to optimize the edge-cloud partitioning of more complex than chain-like DNN structures. For the practical implementation, DADS relies on a modified version of Caffe. Similar to DADS, the industrial effort Auto-Split \cite{banitalebi2021auto}, and $D^3$\cite{zhang2021dynamic} are DAG-based. $D^3$ follows the three-layer (device, edge, cloud) concept of DDNN \cite{teerapittayanon2017distributed}, as well as parallel distribution across edge nodes similar to, \textit{e.g.}, MoDNN \cite{mao2017modnn}.
SPINN \cite{laskaridis2020spinn} is a PyTorch-based framework for distributed inference that leverages concepts of early exit and dynamic splitting, which can accommodate to run-time changes of the environment. The successor of SPINN, DynO \cite{almeida2021dyno}, introduced several optimization targets and bitwidth optimizations.

\section{The Edge-PRUNE framework}
\label{sec:proposed}

\noindent In contrast to the related work, Edge-PRUNE is based on a dataflow computing model, VR-PRUNE \cite{boutellier22vr-prune}, which formalizes concepts such as communication buffer sizing, conditional execution and graph topology. On the practical side, the model enables, \textit{e.g.}, design time analysis for buffer overflow or deadlock. Below, this model of computation is introduced.

\subsection{Model of computation}
\label{ssec:moc}

\noindent In our model of computation \cite{boutellier22vr-prune}, a DNN application is expressed as a directed graph $G=(A,F)$, where nodes $A$ represent computation (\textit{e.g.}, DNN layers), and edges $F$ represent data buffers between nodes. The edges carry data in first-in-first-out (FIFO) order. The connection point between an edge $f \in F$ and a node $a \in A$ is called a \textit{port} $p_a$ such that $f=\mathit{fifo}(p_a)$ and $parent(p_a)=a$. Within edges, data flows in the form of \textit{tokens} that are data packets of pre-defined size. In the machine learning context, tokens equal to tensors, matrices of intermediate features between DNN layers.

In the dataflow model of computation, nodes are called \textit{actors}. Computation in an actor is triggered based on data availability: an actor starts to compute (it \textit{fires}) when all the input edges of that actor have a sufficient number of tokens available -- for each input port of each actor, the \textit{input token rate} indicates the number of tokens required for one firing. Upon firing, the actor consumes a number of tokens (indicated by the token rate) from each input edge, and produces a specific number of tokens to each departing edge of that actor.

The model of computation \cite{boutellier22vr-prune} heeded by Edge-PRUNE has two special features that set it apart from other dataflow models of computation: a) support for variable token rates, and b) the symmetric token rate requirement. Suppose the application graph $G$ contains two actors $a$ and $b$, which are interconnected by edge $f=\mathit{fifo}(p_a)=\mathit{fifo}(p_b)$. The variable token rate feature specifies for each port $p$ the non-negative integer values \textit{upper rate limit} $url(p)$, the \textit{lower rate limit} $lrl(p)$ and the \textit{active token rate} $atr(p)$ such that $lrl(p) \leq atr(p) \leq url(p)$. Both $lrl(p)$ and $url(p)$ are fixed at application design time, whereas $atr(p)$ is allowed to be set before each firing of $parent(p)$. The symmetric token rate requirement, on the other hand, requires that $atr(p_a)$ = $atr(p_b)$ always holds for for each edge $f \in F$.

Each actor belongs to one of the four pre-defined types: \textit{static processing actor} (SPA), \textit{dynamic actor} (DA), \textit{configuration actor} (CA) or \textit{dynamic processing actor} (DPA). DAs, DPAs and CAs may only appear within so-called \textit{dynamic processing (sub)graphs}, DPGs, that encapsulate the variable-token rate behavior of the application. A DPG consists of a CA, two DAs, and any number of DPAs and/or SPAs. The CA sets the current token rate within the DPG, whereas the DAs and DPAs within the DPGs implement the token rate variability with their input and output ports. If the DPGs follow prescribed design rules and patterns \cite{boutellier22vr-prune}, the DPGs are compile-time analyzable for consistency, \textit{i.e.}, absence of deadlock and/or buffer overflow.

\subsection{Transmit and receive FIFOs}

\noindent The FIFO buffer edges, which interconnect dataflow actors have pre-defined \textit{capacity} (maximum number of tokens that each FIFO can hold at any moment), as well as maximum (\textit{url}) and minimum (\textit{lrl}) token rates. Edge-PRUNE features \textit{transmit FIFO} and \textit{receive FIFO} types to introduce necessary infrastructure for distributed computing.

At application initialization, before any application processing has been done, a receive (RX) FIFO blocks and waits for a remote connection from a matching transmit (TX) FIFO. Once all receive FIFOs of the application graph $G$ have successfully established a connection to the respective transmit FIFO, the application dataflow processing begins.

Introduction of TX and RX FIFOs requires no changes to the application graph $G$: the RX and TX FIFOs are automatically inserted by the Edge-PRUNE framework at the stage of code synthesis, when executable code for endpoint and server devices is generated from the dataflow application specification. Due to the automation of this step, the same application graph and actor descriptions can be used for local (single system) and distributed code generation.

\subsection{Edge-PRUNE framework and tools}

\noindent The behavior of each actor is described in a separate source code file; in the current Edge-PRUNE realization, C and OpenCL C language files are accepted. Each actor description has \textit{initialization}, \textit{firing} and \textit{deinitialization} behaviors defined (for OpenCL C only firing behavior).

The Edge-PRUNE framework follows the approach of model-based design and software synthesis (similar to, \textit{e.g.}, \cite{castrillon2011maps}) for specifying the application and the underlying computing infrastructure. In addition to the application graph, Edge-PRUNE also requires an abstraction of the underlying computing platform, which is provided in the form of an undirected \textit{platform graph} that lists the processing units (such as CPU cores and GPUs), and specifies their interconnections. Consequently, also a \textit{mapping file}, which assigns each actor to exactly one processing unit, is required. Unlike the application graph, the platform graph and the mapping file are specific to each computing platform in the distributed system: in each platform-specific mapping file, each actor is defined either for local or remote execution.  

\textbf{Compiler.} The most important software tool related to Edge-PRUNE is the compiler, which requires as input the application graph, actor behavior files, the platform graph and a mapping file. Given this input, the Edge-PRUNE compiler synthesizes a top-level application file, which is later required by the platform-specific compiler (for instance, \texttt{gcc}) to produce the application executable. The Edge-PRUNE compiler streamlines implementation of distributed computing: at minimum, only the mapping file needs to be modified to reflect changes in the distributed scenario.

\textbf{Explorer.} A central research topic in distributed DNN inference \cite{kang2017neurosurgeon, li2018edge} has been design space exploration for endpoint/server DNN partitioning. In contrast to most frameworks, Edge-PRUNE adopts a profiling-based approach: the Edge-PRUNE Explorer tool indexes the $N$ actors of the application graph into an ascending order based on precedence, and generates $N$ mapping file pairs (one for the endpoint device, and one for the server) by shifting the client-server partitioning point actor-by-actor from the inference input towards the inference output. In addition to the mapping files, the explorer also generates client-side and server-side scripts that enable execution-time profiling of all mapping alternatives. If the test data quantity is set appropriately, profiling of the mapping alternatives of a moderate-sized DNN such as SSD-Mobilenet \cite{howard2017mobilenets} can be very well be accomplished within one hour.

\textbf{Analyzer.} The Edge-PRUNE tools include a prototype graph analyzer, which analyzes application graph $G$ consistency against the VR-PRUNE design rules and patterns \cite{boutellier22vr-prune}. 

\subsection{Edge-PRUNE runtime}

\noindent The heterogeneous parallel processing and distributed computing features of Edge-PRUNE have been implemented to a compact C language library, which is compiled with the actor implementations and the Edge-PRUNE compiler-generated top-level application file into an executable, separately for the endpoint device side and for the server side.

The current Edge-PRUNE runtime relies heavily on Linux inter-process communication and networking functionalities. Each actor that has been mapped for execution on a CPU core, is instantiated as a separate thread, and actor data exchange over FIFOs is synchronized by \textit{mutex} primitives. GPU support is deeply in-built within the Edge-PRUNE runtime such that FIFOs interconnecting CPU and GPU mapped actors, transparently to the application programmer, take care of GPU memory management and data transfers.

In addition to the deeply in-built OpenCL C support, which enables efficient use of most GPUs and, \textit{e.g.}, SIMD instructions of Intel and ARM processors, interfacing with various computation accelerators can also be implemented on the application level: leverage of proprietary Intel oneDNN and ARM CL libraries \cite{boutellier22vr-prune} and CUDA devices \cite{boutellier2018design} has been showcased in our previous works.

The transmit and receive FIFOs for distributed computing have been implemented by Linux \textit{sockets} such that each transmit/receive FIFO pair in an application graph receives a dedicated TCP port number. In this prototype implementation data security is delegated to the connection level, \textit{i.e.}, the endpoint devices are expected to form an SSH connection to the edge server prior to establishing the socket-level connections.

\begin{table*}
\caption{Platforms used for experiments}
\label{proctable}
\begin{tabular}{p{1.5cm}p{7.0cm}p{4.1cm}p{3.8cm}}
\hline\noalign{\smallskip}
Tag & CPU & GPU & Operating system\\
\hline 
i7 & Intel Core i7-8650U, 1.9 GHz, 4(8) cores & Intel UHD Graphics 620 & Ubuntu Linux 18.04 \\
\hline 
N2 & 4$\times$ARM Cortex-A73 and 2$\times$ARM Cortex-A53& ARM Mali G-52 & Ubuntu Linux 18.04 \\
\hline 
N270 & Intel Atom N270, 1.6 GHz, single-core & n/a & Ubuntu Linux 16.04 \\
\hline 
\noalign{\smallskip}
\end{tabular}
\end{table*}

\begin{table}
\caption{Network characteristics}
\label{networktable}
\begin{tabular}{p{2.0cm}p{4.5cm}p{1.1cm}}
\hline\noalign{\smallskip}
Tag & Bandwidth and measured throughput & Latency\\
\hline 
N2-i7 Ethernet & 100 Mbits/s; 11.2 MBytes/s measured & 1.49 ms \\
\hline 
N2-i7 WiFi & 16 Mbits/s; 2.3 MBytes/s measured & 2.15 ms \\
\hline 
N270-i7 Ethernet & 100 Mbits/s; 11.2 MBytes/s measured & 1.21 ms \\
\hline 
N270-i7 Wifi & 72.2 Mbits/s; 4.7 MBytes/s measured & 1.22 ms \\
\hline 
\noalign{\smallskip}
\end{tabular}
\end{table}

\section{Experiments}
\label{sec:experiments}

\noindent The experimental evaluation of Edge-PRUNE was conducted using two convolutional neural networks: an image classification network for vehicle image classification \cite{xie2016resource}, and object tracking based on the SSD-Mobilenet object detector \cite{howard2017mobilenets}. The experimental platforms (see Table~\ref{proctable}) cover an Intel Core i7 based system that acts as the edge server, and two lightweight systems that act as endpoint devices: an ARM multicore single board computer ODROID N2 with a Mali G-52 GPU, and an Intel Atom N270 based single-core system. The endpoint devices were connected with the edge server over an Ethernet cable and WiFi, as summarized in Table~\ref{networktable}.

\begin{figure}
\centering
\includegraphics[width=\linewidth]{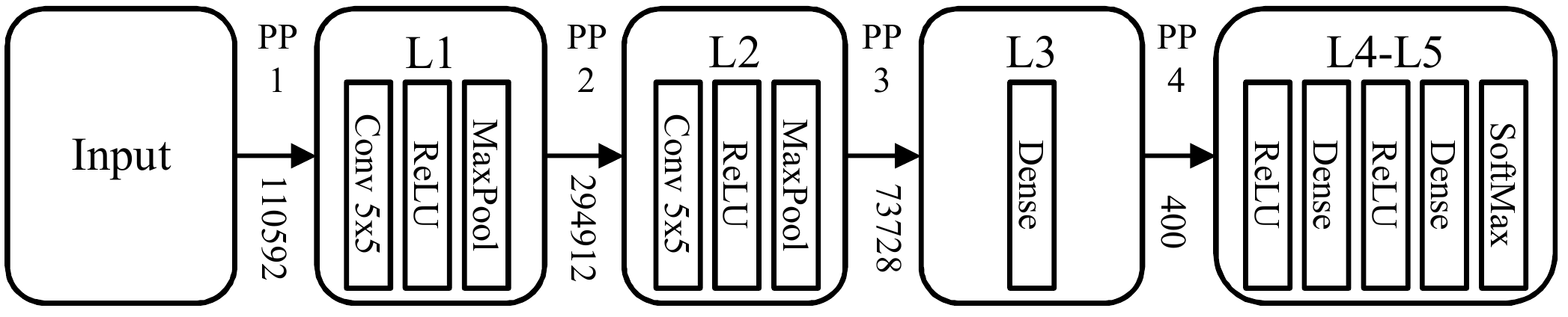}
\caption{The vehicle image classification CNN. Rounded rectangles reflect dataflow actors, and the enclosed smaller rectangles CNN layers. Values between actors indicate edge token size and partition point (PP) index.}
\label{fig:vclass}
\end{figure}

\begin{figure*}
\centering
\includegraphics[width=\linewidth]{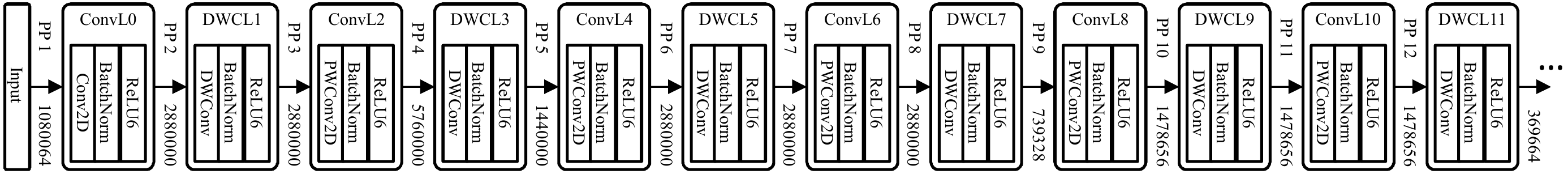}
\caption{The first 13 actors of the SSD-Mobilenet object tracking application. Rounded rectangles are dataflow actors, and the enclosed smaller rectangles are CNN layers. Numeric values indicate edge-specific token size, and PP's are partition points. The entire dataflow graph consists of 53 actors and 69 edges.}
\label{fig:mobilenet}
\end{figure*}

\subsection{Convolutional neural network use cases}
\label{ssec:usecase}
\noindent The CNN for vehicle image classification \cite{xie2016resource} consists of two convolutional layers with 5$\times$5 filter size, max-pooling by a downsampling factor of two, and ReLU activation. The two convolutional layers are followed by three dense layers, of which the first two use ReLU activation, whereas the third dense layer is followed by SoftMax. Fig.~\ref{fig:vclass} details the CNN structure and the mapping of layers to dataflow actors. On the N2 endpoint device, the neural network layer processing was performed by the Mali GPU using ARM Compute Library layer implementations, whereas on the N270 the layers were implemented in plain C language. On the i7 edge server, layer processing of \texttt{L1} and \texttt{L2} actors (Fig.~\ref{fig:vclass}) was performed by the Intel oneDNN library, whereas the computationally simple \texttt{L3} and \texttt{L4-L5} actors were written in plain C language.

The other use case CNN, SSD-Mobilenet (Fig.~\ref{fig:mobilenet}), is a well-known object detector designed especially for mobile applications. Essentially, the CNN architecture is based on the Mobilenet feature extractor \cite{howard2017mobilenets} followed by the single-shot multibox object detector (SSD) \cite{liu2016ssd}. Altogether, SSD-Mobilenet has 129 layers that are grouped into 47 dataflow actors. In addition to the 47 DNN actors, the Edge-PRUNE graph includes 6 actors for non-maximum suppression, object tracking and data I/O. Notably, the SSD-Mobilenet graph is not a straightforward chain, but also contains branches. Both on the N2 endpoint device and on the i7, the DNN layers were executed on the GPU using OpenCL layer implementations.

\begin{figure}
\centering
\includegraphics[width=0.85\linewidth]{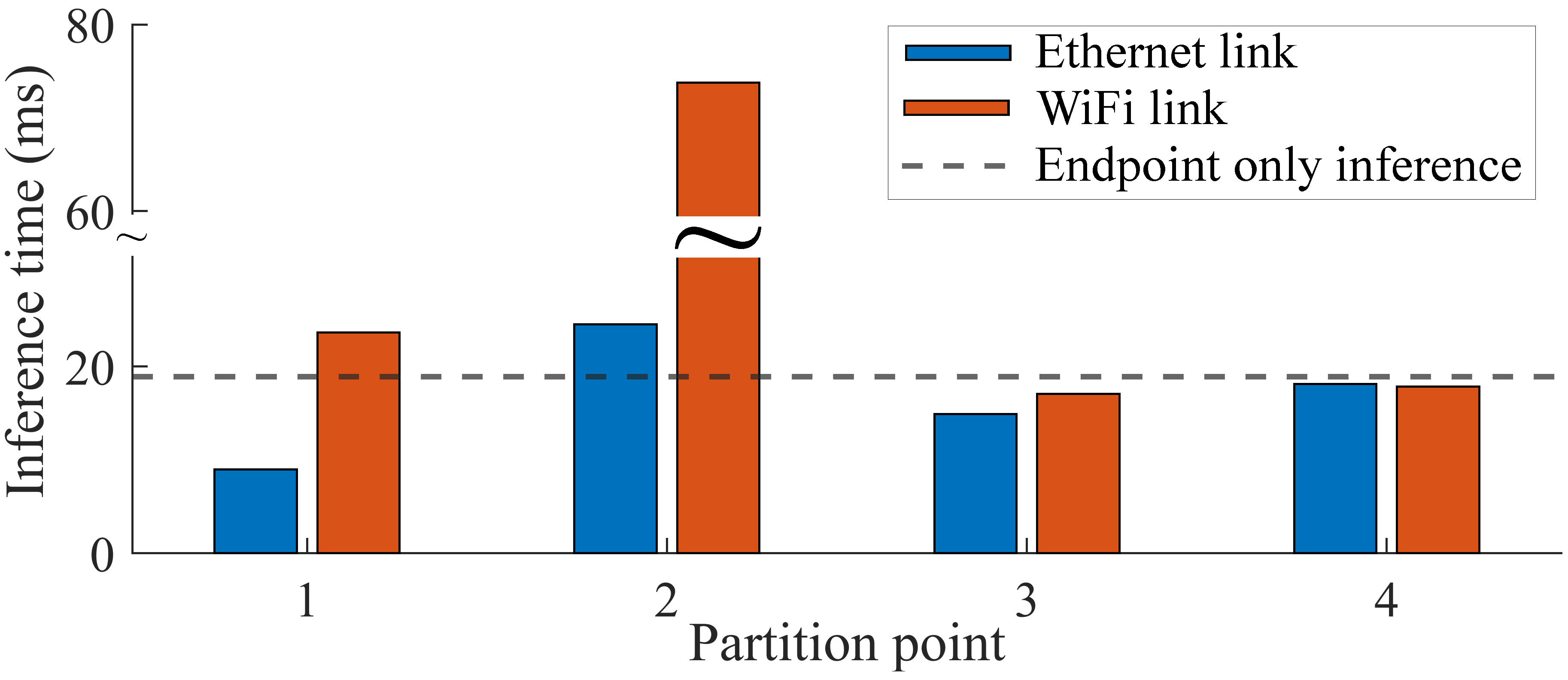}
\caption{Vehicle classification endpoint inference time, when inference is shared between the N2 (end device) and the i7 server, at different partition points.}
\label{fig:odroid-vc}
\end{figure}

\begin{figure}
\centering
\includegraphics[width=0.85\linewidth]{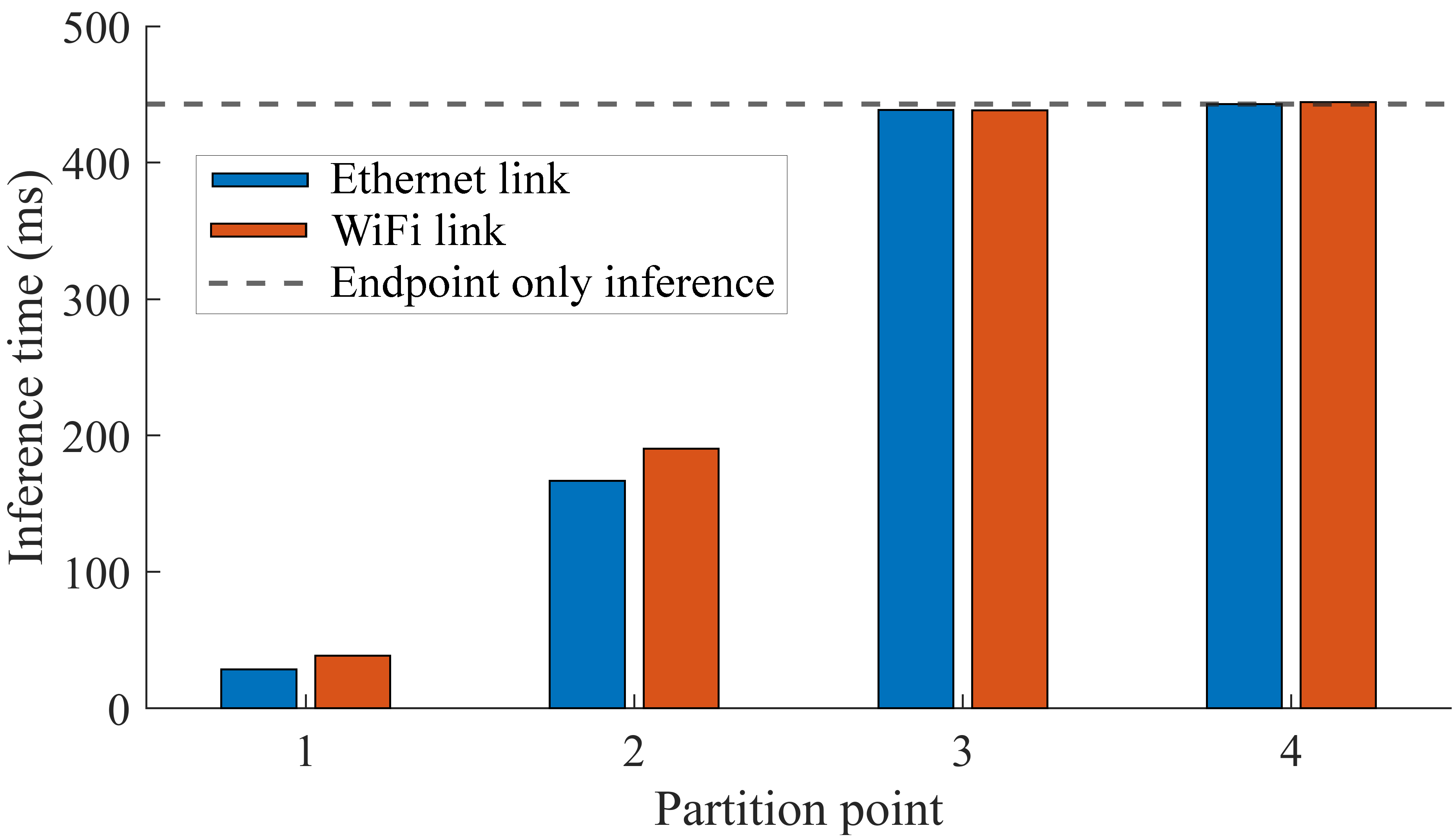}
\caption{Vehicle classification endpoint inference time, when inference is shared between the N270 (end device) and the i7 server, at different partition points.}
\label{fig:atom-vc}
\end{figure}

\subsection{Throughput on image sequences}

\noindent The main type of experiment for Edge-PRUNE was measurement of image classification / object tracking throughput as a function of computation partitioning between the endpoint device and the edge server. In terms of throughput maximization, it is not straightforward to determine how many layers of the neural network should be processed by the endpoint device, and how many should be left to the edge server, as the throughput depends on the endpoint device's computation characteristics, the neural network's intermediate tensor (token) sizes, and network bandwidth.

\textbf{Vehicle image classification on N2-i7.} Fig.~\ref{fig:odroid-vc} shows the average inference time per input image (384 frames total) on the N2 platform for vehicle image classification (Fig.~\ref{fig:vclass}) over 100 Mbit Ethernet and 16 Mbit Wifi at different CNN partition points. Performing the complete inference on the endpoint device takes 18.9 ms per frame (horizontal dashed line in Fig.~\ref{fig:odroid-vc}), whereas transmitting the raw input data from the endpoint device over Ethernet to be processed by the edge server (blue bar at PP 1 in Fig.~\ref{fig:odroid-vc}) would take only 9.0 ms per frame. However, if transmission of raw image data outside the endpoint device is to be avoided due to privacy concerns, the best throughput is achieved when \texttt{Input}, \texttt{L1} and \texttt{L2} actors (Fig.~\ref{fig:vclass}) are processed on the endpoint device, and actors \texttt{L3}, \texttt{L4-L5} are processed by the edge server, which yields 14.9 ms processing time (PP 3 in Fig.~\ref{fig:odroid-vc}) on the endpoint device. If the slower WiFi connection (red bars in Fig.~\ref{fig:odroid-vc}) is used, transmission of raw image data to the edge server becomes slower than full endpoint device inference, whereas the optimal inference partitioning between the endpoint device and edge server remains the same as with the faster Ethernet connection (PP 3 in Fig.~\ref{fig:odroid-vc}), although somewhat slower: 17.1 ms per frame. One of the main reasons why this partition point is optimal on both the slow and fast connection, can be seen in the token sizes (Fig.~\ref{fig:vclass}) between the actors: the token size between \texttt{L2} and \texttt{L3} is 73728 bytes, much less than, \textit{e.g.}, between \texttt{L1} and \texttt{L2} (294912 bytes), which reduces communication time.

\textbf{Vehicle image classification on N270-i7.} For a sequence of 16 images, full inference on the endpoint took 443 ms per frame. Consequently, Fig. \ref{fig:atom-vc} shows that collaborative inference improves inference throughput significantly. If raw input data transmission to the edge server (28.6 ms over Ethernet, 38.9 ms over WiFi) is not possible due to privacy concerns, processing actors \texttt{Input} and \texttt{L1} on the N270, and actors \texttt{L2}, \texttt{L3} and \texttt{L4-L5} on the edge server considerably reduces inference time per frame (Ethernet: 167 ms, WiFi: 191 ms).

\textbf{SSD-Mobilenet object tracking on N2-i7.} This CNN is significantly deeper than the vehicle classification one, and takes 2360 ms for object detection and tracking per input frame (with an image sequence of 10 frames), when inference is done fully on the N2 endpoint device. With this application, the throughput increase by collaborative inference was significant: over Ethernet, by letting the endpoint device perform the inference between actors \texttt{Input} ... \texttt{DWCL9}, and offloading the rest to the edge server, the client's inference time dropped to 406 ms, which is a 5.8$\times$ throughput increase. For the slower WiFi connection, minimal end device inference time was 470 ms at partition point 9.

\subsection{Dual-input vehicle image classification} 
\noindent The vehicle image classification application was also realized as a two-input case, similar to Fig.~\ref{fig:overview}, where actors \texttt{Input} through \texttt{L3} were replicated into two instances each, joining at a two-input \texttt{L4L5} actor. The 1st instances of \texttt{Input}, \texttt{L1}, \texttt{L2} and \texttt{L3} were mapped to the N2 platform, whereas the 2nd instance of \texttt{Input} was mapped to the N270, and the rest of the actors to the i7 edge server. On this configuration, the inference time was 49 ms on the N270, 154 ms on the N2, and 157 ms on the server.  

\subsection{Single-input end-to-end latency}

\noindent Besides throughput, the efficiency of Edge-PRUNE was also evaluated for the case of single image end-to-end inference latency. The vehicle classification DNN application was distributed over the N2 (endpoint) and i7 (edge server) devices such that \texttt{L1} and \texttt{L2} actors were assigned to the N2, and the rest of the actors to the i7. The two systems were interconnected by the 100 Mbit Ethernet cable connection. The vehicle classification Edge-PRUNE application was modified slightly to include a feedback socket connection from the edge server-mapped \texttt{L4-L5} actor back to the endpoint device. Upon completing the inference processing, the edge server sent back a signal over the socket connection to notify the endpoint device of completed inference.

Using this setup, the vehicle classifier application was measured to provide 31.2 ms of end-to-end latency from data input on the endpoint device side, until providing the classification result on the edge server side. Detailed profiling further revealed that 57\% (17.5 ms) of this time was spent on endpoint device inference, 23\% (7.3 ms) in communication over Ethernet, and 20\% (6.3 ms) on edge server inference. It needs to be pointed out that inference time for single images much slower than inference for image sequences  (Fig.~\ref{fig:odroid-vc}) due to CPU cache behavior.

\begin{figure}
\centering
\includegraphics[width=\linewidth]{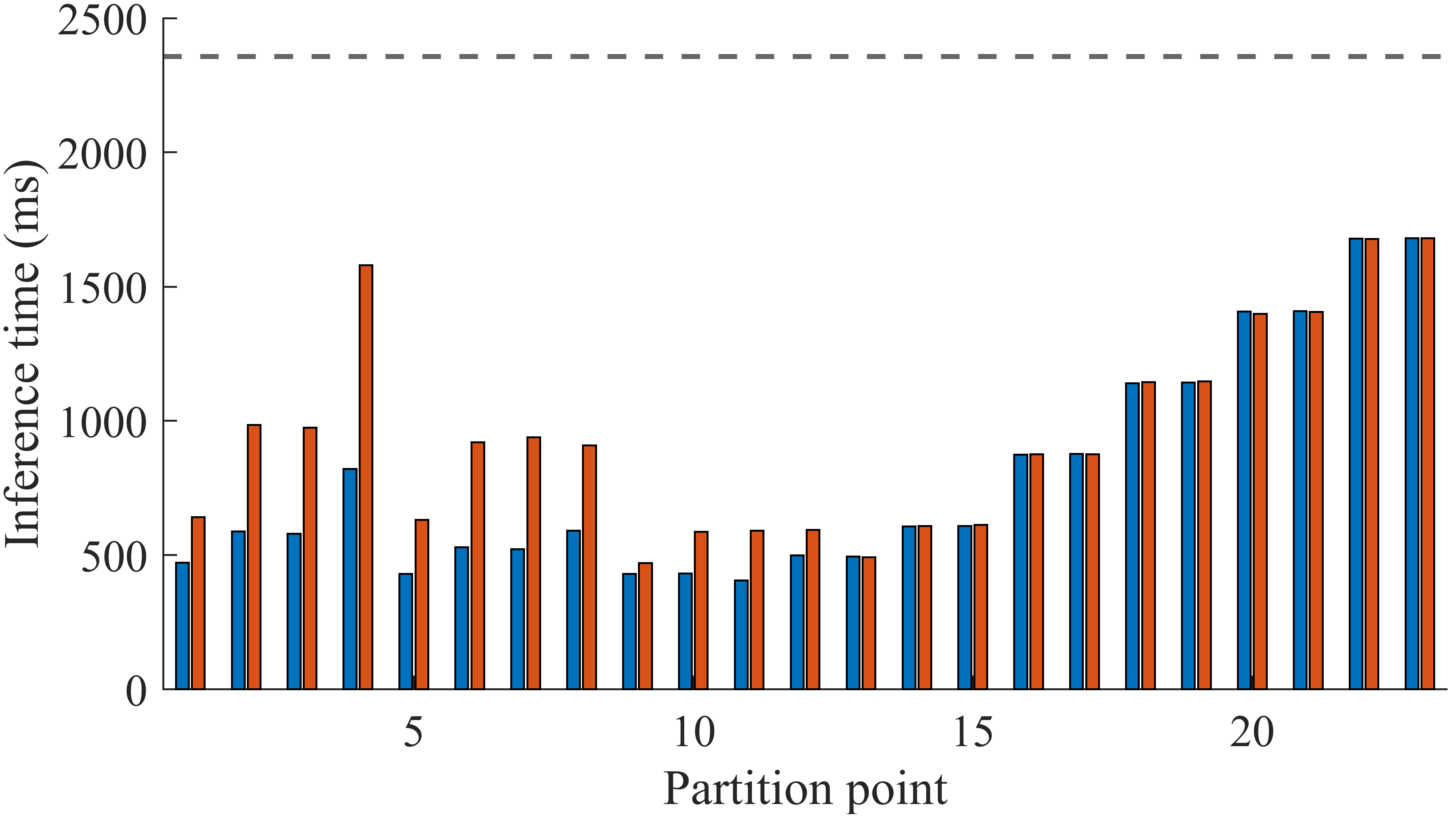}
\caption{Endpoint inference time for SSD-Mobilenet, when inference is shared between the N2 (end device) and the i7 server, at different partition points.}
\label{fig:n2-mobilenet}
\end{figure}

\section{Discussion}
\label{sec:discussion}

\noindent Even though Edge-PRUNE was presented here as a framework for distributed machine learning inference, the framework is equally suitable for distributed processing of other signal processing type workloads such as wireless communications (see, \textit{e.g.}, the previous work \cite{boutellier22vr-prune}). In the same vein, the generic dataflow infrastructure of Edge-PRUNE lends itself also to further actor network topologies such as distributing computation output to more than one server (single-input, multiple output, or multiple-input, multiple-output), although such configurations were not presented in this work.

\section{Conclusions}
\label{sec:conclusions}

\noindent This paper presented Edge-PRUNE, a computation framework for distributed machine learning inference, with key characteristics of formality, flexibility and efficiency. Edge-PRUNE is based on the formal VR-PRUNE dataflow model of computation, which provides a solid basis for computation and data transmission both within and between computation platforms. The flexibility of Edge-PRUNE has been illustrated in the experiments by 1) adopting neural network layer implementations simultaneously from mixed libraries (ARM Compute Library, Intel oneDNN and plain C / OpenCL), and 2) heterogeneous application graph structures, such as two-input image classification. Finally, the run-time efficiency of Edge-PRUNE has been shown by throughput and latency measurements.


\end{document}